\documentclass[conference]{IEEEtran}

\usepackage[cmex10]{amsmath}
\usepackage{cite}

\usepackage{latexsym}
\usepackage{color}
\usepackage{graphics}
\usepackage{amssymb}
\usepackage{amsthm}

\usepackage{cite}
\usepackage{array}
\usepackage{graphicx}

\usepackage{url}

\newcommand{\fracSum}[1]{{\underset{{#1}}{\sum}}}

\newcommand{\vect}[1]{\mathbf{#1}}

\newcommand{\maximize}[1]{{\underset{{#1}}{\mathrm{maximize}}}}

\theoremstyle{plain}
\newtheorem{remark}{Remark}

\newtheorem{theorem}{Theorem}

\newtheorem{lemma}{Lemma}

\newtheorem{proposition}{Proposition}

\begin{document}

\title{Designing Wireless Broadband Access for Energy Efficiency: Are Small Cells the Only Answer?}

\IEEEoverridecommandlockouts

\author{\IEEEauthorblockN{Emil Bj{\"o}rnson\IEEEauthorrefmark{1}, Luca Sanguinetti\IEEEauthorrefmark{2}\IEEEauthorrefmark{3}, and Marios Kountouris\IEEEauthorrefmark{3}\IEEEauthorrefmark{4}}
\IEEEauthorblockA{\IEEEauthorrefmark{1}\small{Department of Electrical Engineering (ISY), Link\"{o}ping University, Link\"{o}ping, Sweden}}
\IEEEauthorblockA{\IEEEauthorrefmark{2}\small{Dipartimento di Ingegneria dell'Informazione, University of Pisa, Italy} $\quad$\IEEEauthorrefmark{3}\small{LANEAS, CentraleSup{\'e}lec, Gif-sur-Yvette, France}}
\IEEEauthorblockA{\IEEEauthorrefmark{4}\small{Mathematical and Algorithmic Sciences Lab, France Research Center, Huawei Technologies Co. Ltd.}
\thanks{This paper was supported in part by ELLIIT,  the People Programme (Marie Curie Actions) FP7 PIEF-GA-2012-330731 Dense4Green, and the ERC Starting Grant 305123 MORE.}%
} \vspace{-4mm}
}

\maketitle

\begin{abstract}
The main usage of cellular networks has changed from voice to data traffic, mostly requested by static users. In this paper, we analyze how a cellular network should be designed to provide such wireless broadband access with maximal energy efficiency (EE). Using stochastic geometry and a detailed power consumption model, we optimize the density of access points (APs), number of antennas and users per AP, and transmission power for maximal EE. Small cells are of course a key technology in this direction, but the analysis shows that the EE improvement of a small-cell network saturates quickly with the AP density and then ``massive MIMO'' techniques can further improve the EE.
\end{abstract}

\IEEEpeerreviewmaketitle

\vspace{-1mm}

\section{Introduction}

The data traffic in cellular networks has experienced an exponential growth in the past couple of decades, and this trend is expected to continue in the foreseeable future \cite{Cisco2014}. Whenever one observes an exponential growth rate, one has to question whether it can be sustained---in particular from an ecological perspective. The energy consumption of the information and communication technology (ICT) industry and its related pollution have already become major societal and economical concerns \cite{Fehske2011a}. To accommodate a $1000 \times$ higher data traffic over the next 10-15 years, without increasing the ICT footprint, we need to design new technologies that improve the \emph{overall energy efficiency} (EE) by $1000 \times$ \cite{Greentouch2013}.

Important steps towards improving the EE of cellular network were taken in \cite{Auer2011a, Tombaz2011a,Bjornson2015a}, where a methodology for measuring the energy consumption was developed. It shows that the radiated signals, radio-frequency (RF) circuits, baseband processing, and backhaul infrastructure are all contributing to the overall energy consumption. A promising way to improve the EE is the \emph{small-cell network} approach \cite{Hoydis2011c}, which is based on deploying a high density of low-power access points (APs) endowed with multiple antennas and advanced beamforming capabilities. Smaller cells require less radiated signal energy since the distances between APs and user equipments (UEs) are shorter. This comes at the price of deploying many more APs, which increase the circuit energy consumption instead \cite{Bjornson2013e}. Another promising approach to improve the EE is \emph{massive MIMO} (multiple-input multiple-output) \cite{Marzetta2010a,Hoydis2013a,Larsson2014a,Bjornson2015a}, where each AP serves tens of UEs in parallel by beamforming from hundreds of small antennas. The form factor of the antenna array is not ``massive'' but may be as a flat-screen television or smaller.
The array gain allows for radiating less energy, but the circuit power per AP increases with the number of antennas---the multiplexing gain may, on the other hand, reduce the circuit power spent per UE. Designing cellular networks for high EE is thus a non-trivial task, where the throughput, AP density and hardware characteristics need to be considered.

The spatial distribution of UEs is highly heterogeneous in practice, which calls for an increasingly heterogeneous and complex AP deployment. A promising way to model and analyze such networks is by using \emph{stochastic geometry}~\cite{Haenggi-2009}, where the AP locations form a realization of a spatial point process, typically a Poisson point process (PPP). This approach can provide tractable expressions for key performance metrics such as the coverage probability and average spectral efficiency (SE) in the network. A few prior works have also derived EE-related performance expressions and showed how these depend on the AP and UE densities; for example, \cite{Cao2012a} considered the deployment of two types of single-antenna APs, while \cite{Li2014a} studied the EE when multi-antenna APs serve one UE each.

In this paper, we consider networks with $M$-antenna APs, each serving $K$ UEs. In contrast to our previous work \cite{Bjornson2015a}, which considered spatially symmetric AP deployments, we use stochastic geometry with a given AP density $\lambda$ to model the spatial randomness in network deployment. We obtain a lower bound on the achievable EE and maximize it analytically with respect to $M$, $K$, $\lambda$, and the transmission power $\rho$. The resulting expressions reveal the fundamental interplay between these four design parameters, which are also illustrated numerically.

\section{System Model}

\begin{figure}
\begin{center}
\includegraphics[width=\columnwidth]{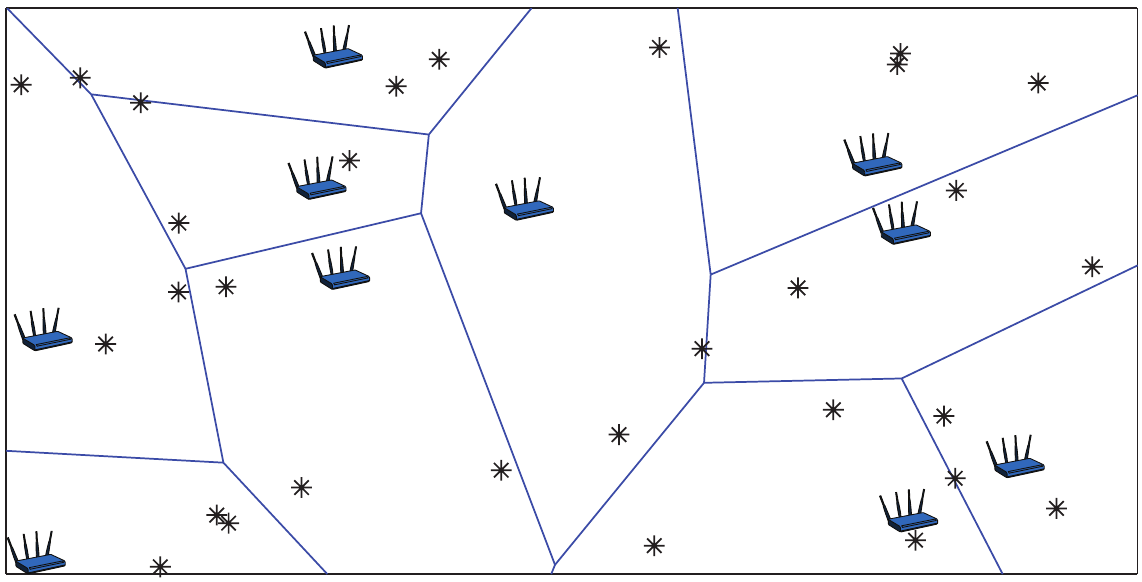}
\end{center}\vskip-4mm
\caption{Illustration of the AP positions modeled by the homogeneous PPP $\Psi_{\lambda}$. Each AP has $M=4$ antennas and serves $K=3$ star-marked UEs.} \label{figureVoronoi} \vskip-4mm
\end{figure}

This paper considers a heterogeneous network with outdoor APs that provide wireless broadband access to stationary UEs in a two-dimensional coverage area. The APs are distributed in $\mathbb{R}^2$ according to a homogeneous PPP $\Psi_{\lambda}$ of intensity $\lambda$ [APs per $\textrm{m}^2$]. Each AP is equipped with $M$ antennas and communicates with $K$ single-antenna UEs, which are uniformly distributed in the Voronoi cell that the AP has as coverage area; see Fig.~\ref{figureVoronoi}. The stationarity of the UEs implies that the coherence interval is sufficiently large so that the overhead required for channel estimation can be reasonably neglected. The channel estimation errors are also neglected as the distortion noise caused by hardware impairments is assumed to be the dominant one \cite{Bjornson2014a}. These assumptions are well-justified in a wireless broadband access scenario where each UE is static (e.g., located in a home) and demands continuously high data traffic (e.g., for video streaming).

We consider the downlink and utilize the translation-invariance of PPPs to concentrate on a \emph{typical UE}, which is representative for any UE in the network \cite{Baccelli2008a}. It is connected to $\mathrm{AP}_0 \in \Psi_{\lambda}$ and has an arbitrary UE index $k$. The received baseband signal $y_k \in \mathbb{C}$ at the typical UE is modeled as
\begin{align} \notag
y_k &= \sqrt{1-\epsilon^2} \left(\sqrt{\rho}  \vect{h}_{0,k}^H \vect{W}_{0} \vect{s}_{0} + \sum_{i \in \Psi_{\lambda} \setminus \{ \mathrm{AP}_0 \}} \sqrt{\rho} \vect{h}_{i,k}^H \vect{W}_{i} \vect{s}_{i} \right) \\ &+ e_k + n_k \label{eq:system-model}
\end{align}
where $\vect{h}_{i,k} \sim \mathcal{CN}(\vect{0}, \omega^{-1} d_{i,k}^{-\alpha} \vect{I}_M)$ is the Rayleigh flat-fading channel from $\mathrm{AP}_i \in \Psi_{\lambda}$ to the typical UE $k$, $d_{i,k}$ is the distance between them, $\alpha> 2$ is the pathloss exponent, and $\omega$ models fixed propagation losses (e.g., wall penetration). We let $\rho$ denote the average RF transmission energy per symbol, $\vect{s}_{i} \sim \mathcal{CN}(\vect{0},\vect{I}_K)$ contains the normalized information symbols sent by $\mathrm{AP}_i$, and define $\vect{W}_{i} \in \mathbb{C}^{M \times K}$ as the corresponding beamforming matrix with normalized columns. The additive receiver noise is modeled by $n_k \sim \mathcal{CN}(0,\sigma^2)$ with $\sigma^2$ being the noise energy per symbol, while $\epsilon$ and $e_k$ model the distortions from hardware impairments and are described next.

\subsection{Distortion Noise from Hardware Impairments}

Contrary to most other works on energy efficiency in wireless networks, the system model in \eqref{eq:system-model} includes distortion noise from hardware impairments. These distortions are due the unavoidable clock drifts in local oscillators, finite-precision analog-to-digital converters, non-linearities, finite-order analog filters, etc. Such impairments cannot be fully removed  \cite{wenk2010mimo}, but are negligible at low signal-to-noise ratio (SNR). However, the impact at high SNR cannot be neglected since it fully characterizes the maximal achievable throughput \cite{Bjornson2013c}.

Similar to \cite{wenk2010mimo,Zhang2012a,Bjornson2013c}, we model the transceiver hardware impairments as reduction of the received signals by a factor $\sqrt{1-\epsilon^2}$ and replacing it with Gaussian distortion noise that carries the removed power: \vspace{-3mm}
\begin{equation} \label{eq:variance-distortion}
e_k \sim \mathcal{CN} \left( \vect{0}, \epsilon^2 \rho  \sum_{i \in \Psi_{\lambda}} \|\vect{h}_{i,k}^H \vect{W}_{i}\|^2 \right),
\end{equation}
where $\| \cdot \|$ is the Euclidean norm. The parameter $0 \leq \epsilon < 1$ is tightly connected to the \emph{error vector magnitude (EVM)}, which is a common measure of the transceiver quality. The EVM is defined as the ratio between average distortion magnitude and the signal magnitude, which becomes $\frac{\epsilon}{\sqrt{1-\epsilon^2}} \approx \epsilon$ with our notation. Typical EVM values are in the range $\epsilon < 0.17$, where smaller EVMs allow for higher SEs \cite{Holma2011a}.

\subsection{Average Achievable Spectral Efficiency}

The SE of a communication link is strictly upper bounded by the channel capacity. In contrast, there is no strict upper bound on the EE metric of a communication link; \cite{Bjornson2014a} showed that one can achieve an unbounded EE in massive MIMO if the circuit power is neglected---and only time will tell how small the circuit power becomes in future hardware. In other words, one can only define achievable lower bounds on the EE of a network. In this paper we derive a lower bound that is tractable for analytic optimization, and for that we need a closed-form SE expression. We consider zero-forcing beamforming (ZFBF) since it gives the highest EE in the single-cell analysis of \cite{Bjornson2015a}. The typical ergodic achievable SE with ZFBF is\footnote{The SE in \eqref{eq:instant-SE} is obtained by canceling intra-cell interference using ZFBF (see  \cite[Sec.~3.4]{Bjornson2013d} for details), treating inter-cell interference as worst-case Gaussian noise in the decoding, and using the distortion noise variance in \eqref{eq:variance-distortion}.} 
\begin{equation}
\mathbb{E} \left\{ \log_2 \! \left( \!1 + \frac{ (1-\epsilon^2) |\vect{h}_{0,k}^H \vect{w}_{0}^{\mathrm{ZF}}|^2 }{ \!\!\!\! \fracSum{i \in \Psi_{\lambda} \setminus \{ \mathrm{AP}_0 \} } \!\!\!\! \|\vect{h}_{i,k}^H \vect{W}_{i}^{\mathrm{ZF}} \|^2 + \epsilon^2 |\vect{h}_{0,k}^H \vect{w}_{0}^{\mathrm{ZF}}|^2  + \frac{\sigma^2}{\rho}}  \!\right) \right\} \!\label{eq:instant-SE}
\end{equation}
where the expectation is with respect to channel fading for a given $\Psi_{\lambda}$. We now take the average of \eqref{eq:instant-SE} over the PPP. This is hard to do in closed form, but we can obtain a lower bound.

\begin{proposition} \label{prop:average-SE}
An achievable lower bound on the average SE  [bit/symbol/user]  of the network with ZFBF ($M \geq K+1$) is 
\begin{equation}
\widetilde{\mathrm{SE}} = \log_2 \! \left( 1 + \frac{ (1-\epsilon^2) (M-K)  }{ \frac{2 K}{\alpha-2} + \epsilon^2 (M-K) + \frac{\Gamma(\alpha/2+1)}{(\pi \lambda)^{\alpha/2}} \frac{\omega \sigma^2}{\rho}}  \right). \label{eq:average-SE}
\end{equation}
\end{proposition}
\begin{IEEEproof}
The proof is based on taking the average of \eqref{eq:instant-SE} with respect to the channel realizations and the PPP $\Psi_{\lambda}$, and then computing an achievable lower bound by using Jensen's inequality as $\mathbb{E}\{ \log_2(1+\frac{1}{x}) \} \geq  \log_2(1+\frac{1}{\mathbb{E}\{ x \}})$.

Computing the expectation with respect to the channel realizations yields $\mathbb{E}\{ \frac{1}{|\vect{h}_{0,k}^H \vect{w}_{0}^{\mathrm{ZF}}|^2} \, | d_{0,k} \} = \frac{\omega d_{0,k}^{\alpha}}{M-K}$ due to the properties of complex Wishart matrices \cite{Maiwald2000a} and $\mathbb{E}\big\{ \|\vect{h}_{i,k}^H \vect{W}_{i}^{\mathrm{ZF}} \|^2 \, | d_{i,k} \big\} = K \omega^{-1} d_{i,k}^{-\alpha}$ for $i \neq 0$ since the ZFBF at $\mathrm{AP}_0$ is independent of channel realizations in other cells.

The distance to the serving BS is $d_{0,k} \sim \mathrm{Rayleigh} \big( \frac{1}{\sqrt{2\pi \lambda}} \big)$ \cite{Weber2010a}.
We keep $d_{0,k}$ fixed and compute the expectation with respect to the interfering APs (which are further away) as $\mathbb{E}\{ \sum_{i \in \Psi_{\lambda} \setminus \{ \mathrm{AP}_0 \} } d_{i,k}^{-\alpha} \, | d_{0,k}  \} = 2\pi \lambda \int_{d_{0,k}}^{\infty} x^{1-\alpha} dx = \frac{2\pi \lambda d_{0,k}^{2-\alpha}}{\alpha-2}$ \cite[Proposition 2.13]{Weber2010a}. Finally, we make use of $\mathbb{E}\{ d_{0,k}^\nu \} =  \frac{\Gamma(\nu/2+1)}{(\pi \lambda)^{\nu/2}}$ for $\nu > - 2$ (e.g., $\nu =2$ and $\nu = \alpha$) to get \eqref{eq:average-SE}.
\end{IEEEproof}

\vspace{1mm}

The lower bound on the average SE expression in Proposition \ref{prop:average-SE} is used in the remainder of this paper to define an achievable EE and to optimize this metric. The tightness of the lower bound in Proposition \ref{prop:average-SE} is evaluated in Section \ref{sec:simulations}.

\section{Metric \& Problem Formulation}
\label{sec:problem-formulation}

The key performance metric in this paper is the achievable downlink \emph{energy efficiency} (EE) [bit/Joule]. It is defined as the ratio between the area spectral efficiency (ASE) [bit/symbol/$\textrm{m}^2$] and the area energy consumption (AEC) [Joule/symbol/$\textrm{m}^2$]. For a given SE the ASE is defined as
\begin{equation}
\mathrm{ASE} = \lambda K \, \mathrm{SE} \label{eq:def-ASE}.
\end{equation}
The \emph{overall} AEC accounts for radiated signal energy, dissipation in circuits, digital signal processing, backhaul signaling, and overhead such as cooling. These are all non-negligible parts of the energy consumption of practical systems \cite{Auer2011a}. We utilize the generic power consumption model from \cite{Bjornson2015a} and let the AEC take the following general form:
\begin{align}
\mathrm{AEC}  &= \lambda \left(   \frac{K \rho}{\eta} + \mathcal{C}_0 + \mathcal{C}_1 K + \mathcal{D}_0 M + \mathcal{D}_1 M K \right)  + \mathcal{A} \, \mathrm{ASE} \label{eq:def-AEC}
\end{align}
\vspace{-4mm}

\noindent where $\eta \in (0,1]$ is the efficiency of the RF amplifiers at the AP and we recall that $\rho$ is the average RF transmission energy per symbol per active UE. The term $\mathcal{C}_0$ is the static energy consumption of an AP, while $\mathcal{C}_1 K$ and  $\mathcal{D}_0 M$ are the terms that scale linearly with the number of active UEs and number of antennas at the AP. The higher-order term $\mathcal{D}_1 M K$ accounts for the cost of signal processing, in particular, computing the product ${\vect{W}}_{i} {\vect{s}}_{i}$ for each information vector ${\vect{s}}_{i}$ and other matrix operations.\footnote{The precoding matrix $\vect{W}_{i}$ needs to be computed once per coherence interval with a complexity proportional to $MK^2$. Since the coherence interval is assumed to be very long herein, we neglect this part in the paper. See \cite{Bjornson2015a} for a full framework that takes also the precoding computation into account.} The energy consumed by the coding and decoding process (and also backhaul signaling) is proportional to the ASE with proportionality coefficient $\mathcal{A}$. More details on this energy consumption model can be found in \cite[Section IV]{Bjornson2015a}. In the analysis, the parameters $\mathcal{A}$, $\eta$, $\mathcal{C}_j$, and $\mathcal{D}_j$ are fixed at arbitrary positive values. Example values are given in Table \ref{table_parameters_hardware}.

In summary, the EE is defined as $\frac{\mathrm{ASE}}{\mathrm{AEC}}$ using \eqref{eq:def-ASE} and \eqref{eq:def-AEC}.

\subsection{Problem Formulation}

We use the lower bound in Proposition~\ref{prop:average-SE} to facilitate analytical optimization of the EE, while numerical optimization is done in Section \ref{sec:simulations}. Optimization of a lower bound guarantees achievable results. The considered lower bound on the EE is \vspace{-1mm}
\begin{align} \label{eq:def-EE}
\!\!\!\widetilde{\mathrm{EE}} \!=\! \frac{ \lambda K \,\widetilde{\mathrm{SE}}}{  \lambda \Big(   \frac{K \rho}{\eta} + \mathcal{C}_0 + \mathcal{C}_1 K + \mathcal{D}_0 M + \mathcal{D}_1 M K  + \mathcal{A} K \,\widetilde{\mathrm{SE}} \Big) }.\!
\end{align}
\vspace{-3mm}

\noindent Note that the AP density $\lambda$ appears as a multiplicative factor in both the numerator and the denominator of \eqref{eq:def-EE} and thus cancels out. However, the EE still depends on $\lambda$ as it also appears in the lower bound $\widetilde{\mathrm{SE}}$ in \eqref{eq:average-SE}.

As seen from \eqref{eq:def-EE}, we can regard the EE as a function $\widetilde{\mathrm{EE}}(\rho,\lambda,M,K)$ of the downlink transmission power per UE ($\rho$), the AP density ($\lambda$), the number of antennas per AP ($M$), and the number of active UEs per AP ($K$). All other system parameters are assumed to be fixed in the analysis. To guarantee reasonable user performance, the following EE optimization problem is considered: \vspace{-2mm}
\begin{equation} \label{eq:main-optimization-problem}
\begin{split}
\maximize{\substack{\rho\geq 0, \,\, 0 \leq \lambda \leq \lambda_{\max}\\ M,K \in \mathbb{Z}_+}} &\quad \widetilde{\mathrm{EE}}(\rho,\lambda,M,K) \\
\mathrm{subject} \,\, \mathrm{to} \,\,\,\,\, & \quad \widetilde{\mathrm{SE}} = \gamma, \quad M \geq K+1,
\end{split}
\end{equation}
\vspace{-3mm}

\noindent where $\gamma$ is an SE level that is guaranteed on average to UEs in the network.\footnote{By removing interference and noise, it can be shown that \eqref{eq:main-optimization-problem} is only feasible for $0 \leq \gamma < -2 \log_2(\epsilon )$, thus we assume that $\gamma$ lies in this interval.} Note that $\rho$ can be any positive number, while $\lambda$ is a positive number upper bounded by $\lambda_{\max}$; this is the highest average AP density that can be physically deployed. $M$ and $K$ belong to the set $\mathbb{Z}_+$ of strictly positive integers. The EE optimization problem in \eqref{eq:main-optimization-problem} is solved in the next section.

\vspace{-1mm}

\section{Optimizing the Energy Efficiency}
\label{sec:optimization}

In this section, we solve the EE optimization problem in \eqref{eq:main-optimization-problem}. While doing so, we will also derive expressions that reveal the fundamental interplay between the design parameters.

\vspace{-1mm}

\subsection{Optimizing the AP Density}

We begin the optimization by considering the AP density $\lambda$ (in $\mathrm{AP/m}^2$), when the other parameters are fixed.

\begin{theorem} \label{th:optimal-lambda}
Consider problem \eqref{eq:main-optimization-problem} for given values on $M$, $K$, and $\rho$. If the problem is feasible, the EE metric is monotonically increasing in $\lambda$ and thus maximized at $\lambda^* = \lambda_{\max}$.
\end{theorem}
\begin{IEEEproof}
The AP density $\lambda$ only appears in \eqref{eq:main-optimization-problem} as part of the SE expression in \eqref{eq:average-SE}. The EE is an increasing function of the SE, which is in turn monotonically increasing in $\lambda$. Hence, the EE is maximized at its highest value $\lambda_{\max}$.
\end{IEEEproof}

This theorem validates the intuition that from an EE perspective it is preferable to have as high AP density as possible (since the signal and interference power increase accordingly). We should keep in mind the assumption that every AP has at least one UE. Hence, we want to make the cells as small as possible while keeping all APs active. Even under future very high user densities the inter-AP distances will be at the order of meters, thus making the maximal average AP density $\lambda_{\max}$ a finite number (e.g., $0.1$~AP/$\mathrm{m}^2$). Moreover, this paper uses a non-line-of-sight channel model, which might not be valid at very short distances. Hence, we treat $\lambda$ as a fixed parameter in the range $0 < \lambda \leq \lambda_{\max} < \infty$ in the remainder of the paper.

\begin{remark}
Observe that even if we let $\lambda \rightarrow \infty$ (and consequently let $\rho \rightarrow 0$), the EE has the finite upper limit \vspace{-2mm}
\begin{equation} \label{eq:EE-limit-lambda}
\frac{ K \gamma}{ \Big(  \mathcal{C}_0 + \mathcal{C}_1 K + \mathcal{D}_0 M + \mathcal{D}_1 M K  + \mathcal{A} K \gamma \Big) }
\end{equation} \vspace{-3mm}

\noindent because the transmission power term goes away in the EE expression, while the circuit power consumption remains. Hence, smaller cells will only bring EE improvements till the point where the transmission power becomes negligible and then higher cell density only brings marginal improvements.
\end{remark}

\subsection{Optimizing the Transmission Power}

Next, we find the optimal transmission power per UE: $\rho^*$. The following theorem shows that the SE constraint in \eqref{eq:main-optimization-problem} can be eliminated by selecting $\rho$ appropriately.

\begin{theorem} \label{th:optimal-rho}
For any values on $\lambda$, $M$, and $K$, the SE constraint in \eqref{eq:main-optimization-problem} is satisfied by \vspace{-3mm}
\begin{equation} \label{eq:rho-optimal}
\rho^* =  \frac{ \frac{2^\gamma - 1}{1-2^\gamma \epsilon^2} \frac{ \omega \sigma^2 \Gamma(\alpha/2+1)}{(\pi \lambda)^{\alpha/2}}
}{  M-K - \frac{2^\gamma - 1}{1-2^\gamma \epsilon^2} \frac{2 K}{\alpha-2} }
\end{equation}
if problem \eqref{eq:main-optimization-problem}  is feasible. The problem is infeasible whenever $\rho^*$ is negative (i.e., when the denominator of \eqref{eq:rho-optimal} is negative).
\end{theorem}
\begin{IEEEproof}
Follows by solving the SE constraint for $\rho$.
\end{IEEEproof}

This theorem gives the relationship between $\rho^*$ and other system parameters. The optimal transmission power is inversely proportional to the AP density as $\lambda^{-\alpha/2}$, due to shorter pathlosses when $\lambda$ increases. It is a decreasing function of the number of antennas, $M$, due to the array gain from coherent beamforming; the relationship is as $M^{-1}$ when $M$ is large. Finally, $\rho^*$  increases with $K$ since $K$ makes the denominator of \eqref{eq:rho-optimal} smaller; this is explained by the will to operate at higher SNRs when the inter-cell interference grows stronger.

By substituting $\rho^*$ from Theorem \ref{th:optimal-rho} into \eqref{eq:main-optimization-problem} and taking $\lambda$ as a constant, the EE optimization problem is reduced to
\begin{align} \notag
\maximize{M,K \in \mathbb{Z}_+} &\,\,\, \frac{ K \gamma}{ \Big(   \frac{K \rho^*}{\eta} + \mathcal{C}_0 + \mathcal{C}_1 K + \mathcal{D}_0 M + \mathcal{D}_1 M K   + \mathcal{A} K \gamma \Big) } \\
\mathrm{subject} \,\, \mathrm{to} \,\,\, & \,\,\, (M-K) \geq \frac{(2^\gamma - 1)}{(1-2^\gamma \epsilon^2)} \frac{2 K}{\alpha-2}. \label{eq:main-optimization-problem-modified}
\end{align}

\subsection{Optimizing the Number of AP Antennas}

Next, we find the optimal number of AP antennas, $M$, when $\rho^*$  is given and the other system parameters are fixed.

\begin{theorem} \label{th:optimal-M}
For any given values on $\lambda$ and $K$, the EE metric in \eqref{eq:main-optimization-problem-modified} is maximized by
\begin{equation} \label{eq:M-optimal}
M^* \! = K + \frac{2 K (2^\gamma - 1)}{(\alpha\!-\!2)(1\!-\!2^\gamma \epsilon^2)} +  \sqrt{ \frac{2^\gamma \!-\! 1}{1\!-\!2^\gamma \epsilon^2} \frac{K \eta^{-1}\omega \sigma^2 \Gamma( \frac{\alpha}{2} \!+\! 1)}{ (\pi \lambda)^{\frac{\alpha}{2}} (\mathcal{D}_0 \!+\! \mathcal{D}_1 K) }  }.
\end{equation}
If $M^*$ is not an integer, then the optimum is attained at either the closest smaller or larger integer.
\end{theorem}
\begin{IEEEproof}
Maximizing the EE metric in \eqref{eq:main-optimization-problem-modified} with respect to a real-valued $M$ is identical to minimizing the denominator. This is a convex problem and \eqref{eq:M-optimal} is the point where the first derivative is zero. The convexity implies that the optimal integer $M$ is one of the two closest integers to $M^*$ \cite{Bjornson2015a}.
\end{IEEEproof}

This theorem shows how the optimal number of antennas per AP depends on the other system parameters. In particular, we see that $M^*$ increases roughly linearly with the number of UEs. In contrast, fewer antennas should be used if the AP density increases (i.e., smaller cells) or when the circuit power parameters $\mathcal{D}_0$ and $\mathcal{D}_1$ grow so that it becomes more costly to have additional antennas turned on. Finally, more antennas are needed when the SE constraint $\gamma$ is increased.

\subsection{Optimizing the Number of UEs}

Finally, we find the optimal number of active UEs per AP. Since we have the constraint $M \geq K+1$ in \eqref{eq:main-optimization-problem}, a tractable optimization of $K$ requires that also $M$ is changed. Hence, we let $\beta = \frac{M}{K}$ be fixed and optimize $M$ and $K$ jointly.

\begin{theorem} \label{th:optimal-K}
For any given values on $\lambda$ and $\beta = \frac{M}{K} > 1$, the EE metric in \eqref{eq:main-optimization-problem-modified} is maximized by
\begin{align} \label{eq:K-optimal}
K^* = \sqrt{
\frac{ \frac{2^\gamma - 1}{1-2^\gamma \epsilon^2} \frac{ \omega \sigma^2 \Gamma(\alpha/2+1)}{\eta (\pi \lambda)^{\alpha/2}}
}{ \beta \mathcal{D}_1 ( \beta-1 - \frac{2^\gamma - 1}{1-2^\gamma \epsilon^2} \frac{2}{\alpha-2} )}  + \frac{\mathcal{C}_0}{\beta \mathcal{D}_1} }.
\end{align}
If $K^*$ is not an integer, then the optimum is attained at either the closest smaller or larger integer.
\end{theorem}
\begin{IEEEproof}
Maximizing the EE metric in \eqref{eq:main-optimization-problem-modified} with respect to a real-valued $K$ is identical to minimizing $\widetilde{\mathrm{EE}}^{-1}$, which is a convex problem.
Finding the $K\geq 0$ where the first derivative is zero gives a quadratic polynomial equation, from which \eqref{eq:K-optimal} is the positive root. The convexity implies that the optimal integer-valued $K$ is one of the two closest integers.
\end{IEEEproof}

This theorem shows how the number of active UEs depends on the other system parameters. We notice that $K^*$ is an increasing function of the static energy consumption $\mathcal{C}_0$ and a decreasing function of the circuit coefficient $\mathcal{D}_1$ that represents the power consumed by signal processing (which behaves as $\mathcal{D}_1 MK$). Interestingly, $\mathcal{C}_1$ and $\mathcal{D}_0$ have no impact on $K^*$. Higher cell density (i.e., small cells) implies fewer UEs.

\subsection{Alternating Optimization}
\label{subsec:alternating-optimization}

To summarize, we first noted in Theorem \ref{th:optimal-lambda} that the AP density  $\lambda$ should be as large as it physically and practically can. Next, we eliminated the transmission power $\rho$ from by finding its optimal value in Theorem \ref{th:optimal-rho}. Then, Theorems \ref{th:optimal-M} and \ref{th:optimal-K} showed how to optimize the EE separately with respect to $M$ and $K$. However, we would like to solve the original problem \eqref{eq:main-optimization-problem} \emph{jointly} with respect to all the parameters. We propose the following alternating optimization algorithm:

\begin{enumerate}
\item Fix $\lambda$ at its largest possible value (e.g., $\lambda_{\max}$).
\item Assume that an initial feasible point $(\rho,M,K)$ is given;
\item Optimize  $K$ by using Theorem \ref{th:optimal-K} (and update $\rho^*$);
\item Optimize  $M$ by using Theorem \ref{th:optimal-M} (and update $\rho^*$);
\item Repeat 3) -- 4) until convergence is achieved.
\end{enumerate}

This algorithm converges since the EE has a finite upper bound and is non-decreasing in each iteration. The mixed-integer nature of \eqref{eq:main-optimization-problem} makes it hard to further assess the convergence, but we have the following result.

\begin{lemma} \label{lemma:convergence}
Suppose \eqref{eq:main-optimization-problem} is relaxed by treating $M$ and $K$ as real-valued. The alternating optimization algorithm converges to the global optimum of this relaxed EE optimization problem.
\end{lemma}
\begin{IEEEproof}
By treating $K$ and $\beta = \frac{M}{K}$ as the real-valued optimization variables, it is easy to show that $\widetilde{\mathrm{EE}}^{-1}$ in \eqref{eq:main-optimization-problem-modified} is a jointly convex optimization problem. It then follows from \cite[Prop.~6]{Grippo2000a} that the alternating optimization algorithm converges to the global optimum of the relaxed EE problem.
\end{IEEEproof}

\section{Simulation Results}
\label{sec:simulations}

In this section, we provide numerical results that illustrate the behaviors that were derived analytically in Section \ref{sec:optimization}. A number of hardware and propagation parameters appear in the  ASE and AEC models, and these need to be selected in the simulations. The parameter values are summarized in Table \ref{table_parameters_hardware} and builds upon a similar list of parameter values in \cite{Bjornson2015a}.

\begin{table}[!t]
\renewcommand{\arraystretch}{1.3}
\caption{Simulation Parameters}
\label{table_parameters_hardware} \vskip-2mm
\centering
\begin{tabular}{|c|c|c|}
\hline
\bfseries Parameter & \bfseries Symbol & \bfseries Value \\
\hline

Pathloss exponent & $\alpha$ & $3.76$ \\

Fixed propagation loss & $\omega$ & $35 \, \mathrm{dB}$ \\

Power amplifier efficiency & $\eta$ & $0.39$ \\

Level of hardware impairments & $\epsilon$ & 0.05\\

Symbol time & $S$ & $\frac{1}{2 \cdot 10^7} \,\, \mathrm{[s/symbol]}$ \\

Coding/decoding/backhaul & $\mathcal{A}$ & $1.15 \, \mathrm{[J/Gbit]}$ \\

Static energy consumption & $\mathcal{C}_0$ & $10 \, \mathrm{W} \cdot S \,\, \mathrm{[J/symbol]}$ \\

Circuit energy per active UE &  $\mathcal{C}_1$ & $0.1 \, \mathrm{W} \cdot S \,\, \mathrm{[J/symbol]}$ \\

Circuit energy per AP antenna & $\mathcal{D}_0$ & $1 \, \mathrm{W} \cdot S \,\, \mathrm{[J/symbol]}$ \\

Signal processing coefficient & $\mathcal{D}_1$ & $1.56 \cdot 10^{-10} \, \mathrm{[J/symbol]}$ \\

Noise variance & $\sigma^2$ & $10^{-20} \, \mathrm{[J/symbol]}$ \\

\hline
\end{tabular} \vskip-1mm
\end{table}

\subsection{Optimizing Energy Efficiency}

\begin{figure}
\begin{center}
\includegraphics[width=\columnwidth]{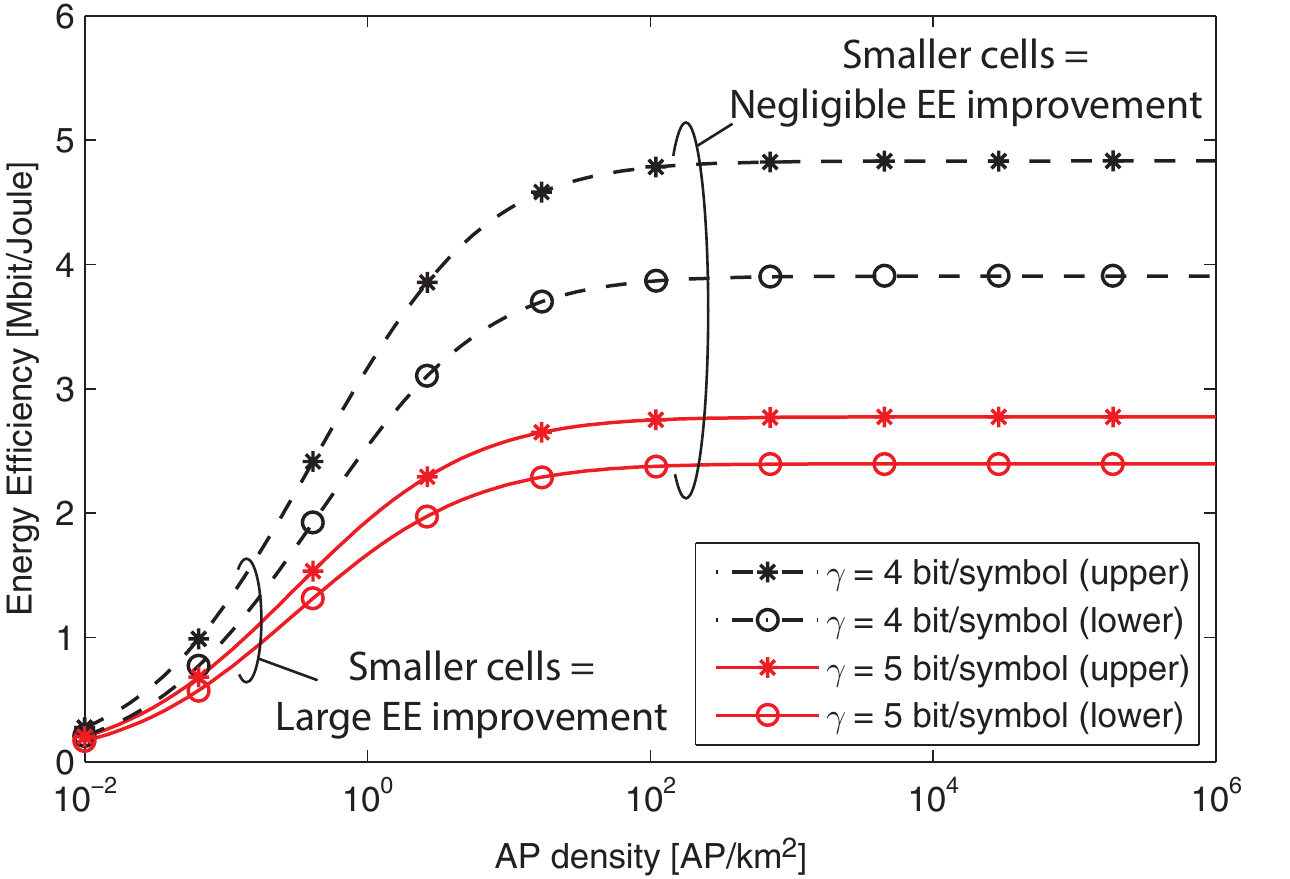}
\end{center}\vskip-5mm
\caption{Energy efficiency (in Mbit/Joule) as a function of the AP density in $\mathrm{AP}/\mathrm{km}^2$, for different SE constraints. The number of antennas, number of UEs, and transmission power are optimized to yield maximum EE.} \label{figureAPdensity} \vskip-4mm
\end{figure}

There are four variables in the EE optimization problem of \eqref{eq:main-optimization-problem}, but we noticed that the AP density $\lambda$ should be as large as it can practically be. We therefore study the EE as a function of $\lambda$.  Fig.~\ref{figureAPdensity} shows the EE for an optimal choice of $M$, $K$, and $\rho$, for different average SE constraints: $\gamma \in \{  4, \, 5 \}$ bit/symbol. Both the lower bound on the SE in Proposition \ref{prop:average-SE} and an upper bound based on Monte-Carlo simulations are shown. There is a gap between these curves, but the behavior is the same---it is basically only a scaling difference.
As expected, increasing the AP density will increase the EE monotonically since the propagation losses reduce. However, when the AP density goes beyond $\lambda = 10^{-4} \, \mathrm{AP/m}^2 = 10^{2} \, \mathrm{AP/km}^2$ the improvements are really minor. This is because the transmission power is now negligible as compared to the circuit power. An AP density of $\lambda = 10^{-4} \, \mathrm{AP/m}^2$ corresponds, roughly speaking, to an average inter-AP distance of 100 meters, which is small but not remarkably small. The behavior is the same for both SE constraints, but 4 bit/symbol gives the highest EE values.

\begin{figure}
\begin{center}
\includegraphics[width=\columnwidth]{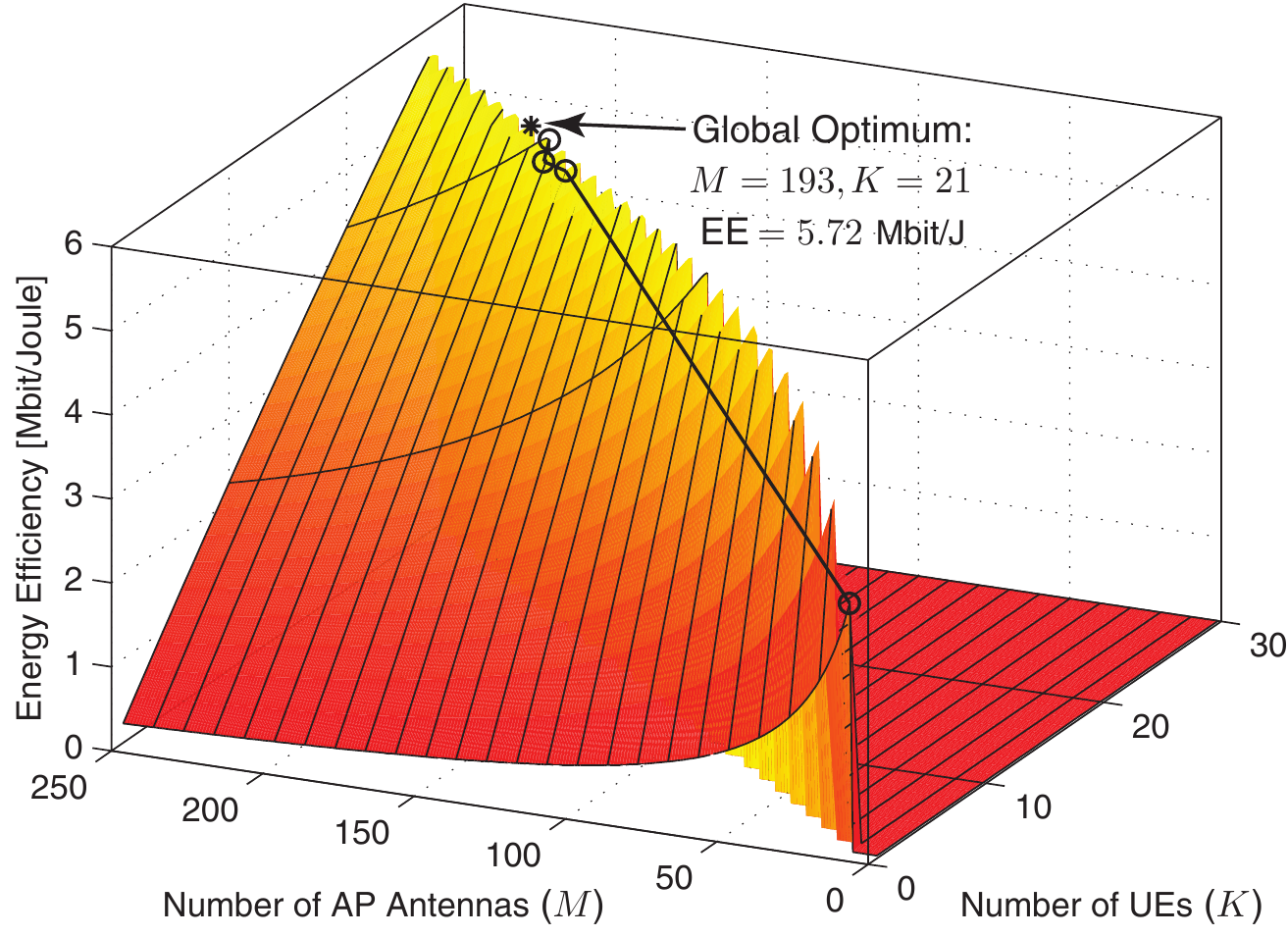}
\end{center}\vskip-4mm
\caption{Energy efficiency (in Mbit/Joule) for $\gamma \!=\! 3$ and $\lambda \!=\! 10^{-4} \,\mathrm{AP}/\mathrm{m}^2 = 10^{2} \, \mathrm{AP}/\mathrm{km}^2$. The global optimum is star-marked, while the convergence
of the alternating algorithm from Section \ref{subsec:alternating-optimization} is indicated with circles.} \label{figure3D} \vskip-4mm
\end{figure}

Next, we fix the AP density to $\lambda = 10^{-4} \, \mathrm{AP/m}^2$ and the SE to 3 bit/symbol, while studying the impact of other parameters. Fig.~\ref{figure3D} shows the EE lower bound as a function of the number of AP antennas ($M$) and number of UEs ($K$), for optimized transmission powers. The strong inter-cell interference creates a need to amplify the desired signals by an array gain from coherent beamforming; many more antennas than UEs are required to support the prescribed SE. The global EE maximum is achieved by $(M^*,K^*) = (193,21)$, and a total transmission power of $\frac{K^* \rho^*}{S} = 424 \,\, \mathrm{mW}$. This type of configurations is known as massive MIMO \cite{Marzetta2010a,Hoydis2013a,Larsson2014a,Bjornson2015a}.

The alternating optimization algorithm from Section \ref{subsec:alternating-optimization} is also considered in Fig.~\ref{figure3D}. It was initiated at $(M,K) = (10,1)$ and converged in three iterations to $(M,K) = (183,20)$ with an EE of $5.71$ Mbit/Joule. The 0.2\% deviation from the global optimum is due to rounding effects, since optimal converge is only guaranteed for real-valued $M$ and $K$ (see Lemma \ref{lemma:convergence}).

\subsection{Optimization under Fixed UE Density}

Suppose the APs are deployed to match a certain density of UEs: $\mu$ UEs per $\mathrm{m}^2$. To accommodate all these UEs, the network should be designed under the additional constraint
\begin{equation} \label{eq:UE-density-constraint}
\mu = K \lambda.
\end{equation}
We now study how this UE density determines the AP density. Future densities from $10^2$ UEs per $\mathrm{km}^2$ (in rural areas) to $10^5$ UEs per $\mathrm{km}^2$ (in shopping malls) have been predicted in the METIS project \cite{METIS_D11_short}, and are used as reference points herein.

\begin{figure}[th!]
\begin{center}
\includegraphics[width=\columnwidth]{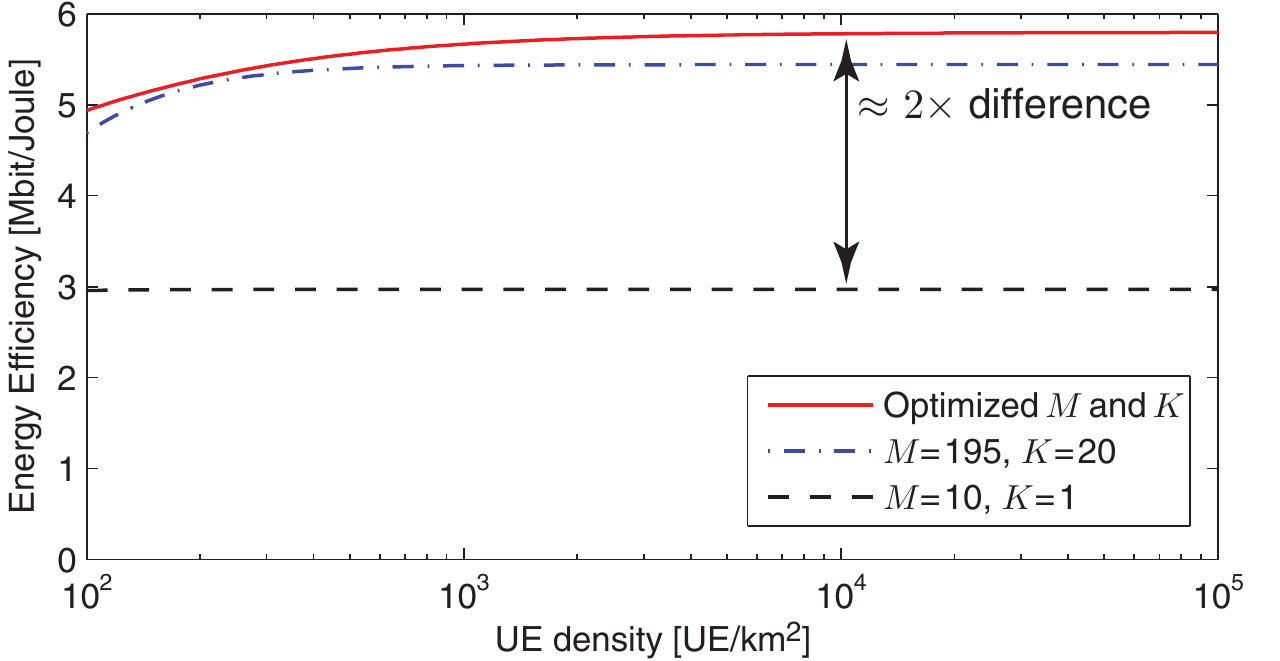}
\end{center}\vskip-4mm
\caption{Energy efficiency (in Mbit/Joule) as a function of the UE density $\mu = K \lambda$. The EE is optimized with respect to $(M,K,\lambda,\rho)$, or only with respect to $(\lambda,\rho)$ for given $M$ and $K$.} \label{figureUserDensityEE} \vskip-4mm
\end{figure}

\begin{figure}[t!]
\begin{center}
\includegraphics[width=\columnwidth]{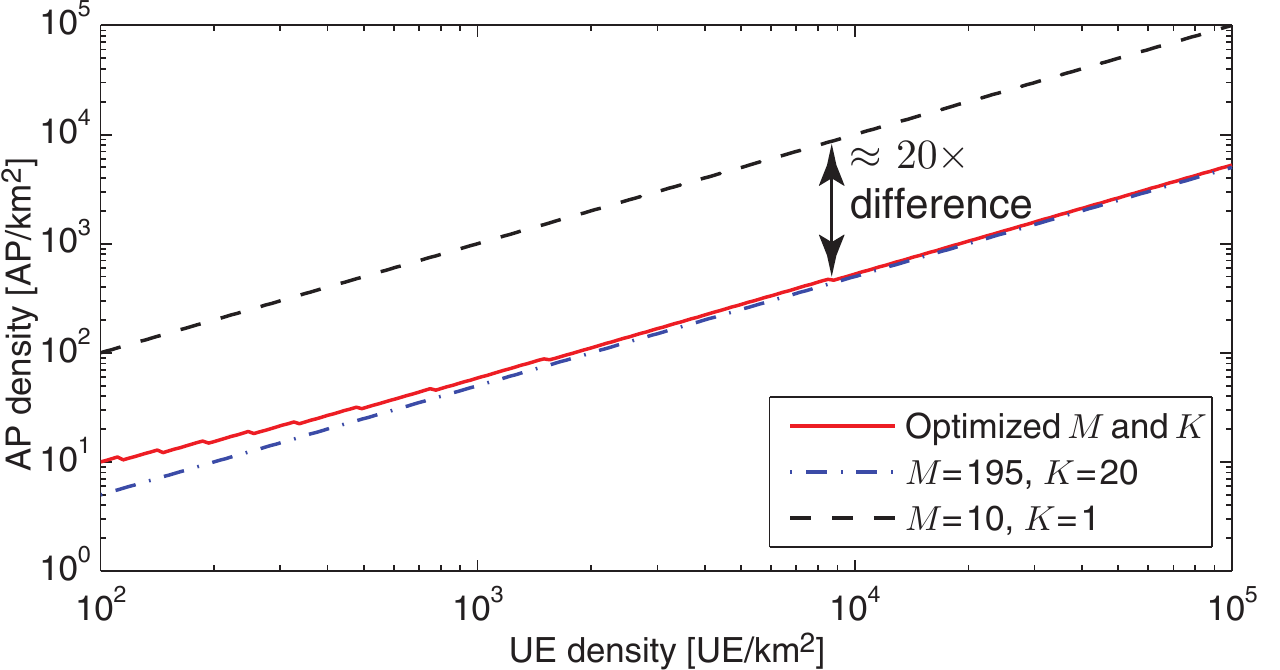}
\end{center}\vskip-4mm
\caption{Optimized AP density (in $\mathrm{AP/km}^2$) as a function of the UE density $\mu = K \lambda$. The system is optimized in the same way as in Fig.~\ref{figureUserDensityEE}.} \label{figureUserAPDensity} \vskip-4mm
\end{figure}

Fig.~\ref{figureUserDensityEE} shows the EE has a function of the UE density for $\gamma = 3$ bit/symbol, while Fig.~\ref{figureUserAPDensity} shows the corresponding AP density. The design parameters $M$, $K$, $\lambda$, and $\rho$ are optimized as in \eqref{eq:main-optimization-problem} but with the additional constraint in \eqref{eq:UE-density-constraint}. Two reference cases are also shown: Single-user transmission with $(M,K) \!=\! (10,1)$; and massive multi-user MIMO transmission with $(M,K) \!=\! (195,20)$. Only the AP density and transmission power were optimized for EE in the two reference cases.

Several important observations can be made. Firstly, almost the same EE can be maintained irrespective of the UE density. This is mainly achieved by scaling the AP density linearly with the UE density, while basically the same number of AP antennas and UEs per AP are used. Secondly, the fixed massive MIMO configuration $(M,K) = (195,20)$ achieves nearly the optimized EE, which is why we chose this reference case. In contrast, single-user transmission in each cell leads to much lower EE. We stress that the optimal configuration (and the fixed massive MIMO case) serves the UEs with the same SE of 3 bit/symbol but using $20 \times$ lower AP density---this is likely to be a key property for cost-efficient AP deployments.

\section{Conclusions}

We have designed cellular networks for energy efficiency. This was formulated as an optimization problem by using stochastic geometry, a lower bound on the SE, and a state-of-the-art power consumption model. The variables were AP density, number of antennas and UEs per AP, and the transmission power. The results show that the EE increases with the AP density, but the positive effect saturates when the circuit power dominates over the transmission power. A further leap in EE is achieved by enabling massive MIMO transmission, by having many AP antennas and spatially multiplex many UEs/cell. The gains come from intra-cell interference suppression and by sharing the circuit power cost between these UEs. The analysis focused on providing wireless broadband access to static users, while future work will consider fast-fading channels where the channel estimation overhead and errors cannot be neglected.

\vspace{-0.5mm}

\bibliographystyle{IEEEbib}
\bibliography{IEEEabrv,refs}

\end{document}